\documentclass[sn-mathphys,pdflatex,iicol]{sn-jnl}

\jyear{2021}%

\theoremstyle{thmstyleone}%
\theoremstyle{thmstyletwo}%

\theoremstyle{thmstylethree}%
\newtheorem{definition}{Definition}%

\begin{document}

\title[Gravitational wave of the Bianchi VII universe: particle trajectories, geodesic deviation and tidal accelerations]{Gravitational wave of the Bianchi VII universe: particle trajectories, geodesic deviation and tidal accelerations}



\author*[1,2]{\fnm{Konstantin} \sur{Osetrin}}\email{osetrin@tspu.edu.ru}

\author[1]{\fnm{Evgeny} \sur{Osetrin}}\email{evgeny.osetrin@tspu.edu.ru}
\equalcont{These authors contributed equally to this work.}

\author[1]{\fnm{Elena} \sur{Osetrina}}\email{elena.osetrina@tspu.edu.ru}
\equalcont{These authors contributed equally to this work.}

\affil*[1]{\orgdiv{Center for Mathematical and Computer Physics}, \orgname{Tomsk State Pedagogical University}, \orgaddress{\street{Kievskaya str. 60}, \city{Tomsk}, \postcode{634061}, 
\country{Russia}}}

\affil[2]{
\orgname{National Research Tomsk State University}, \orgaddress{\street{Lenina pr. 36}, \city{Tomsk}, \postcode{634050}, 
\country{Russia}}}



\abstract{
For the gravitational wave model based on the type III Shapovalov wave space-time, test particle trajectories and the exact solution of geodesic deviation equations for the Bianchi type VII universe are obtained.
Based on the found 4-vector of deviation, tidal accelerations in a gravitational wave are calculated.
For the obtained solution in a privileged coordinate system, an explicit form of transformations into a synchronous reference system is found, which allows time synchronization at any points of space-time with separation of time and spatial coordinates.
The synchronous reference system used is associated with a freely falling observer on the base geodesic.
In a synchronous coordinate system, an explicit form of the gravitational wave metric, a 4-vector of geodesic deviation, and a 4-vector of tidal accelerations in a gravitational wave are obtained. 
The exact solution describes a variant of the primordial gravitational wave.
The results of the work can be used to study the plasma radiation generated by tidal accelerations of a gravitational wave.
}

\keywords{
gravitation wave, cosmology, deviation of geodesics, tidal acceleration, Hamilton-Jacobi equation, Bianchi universes, Stäckel spaces, Shapovalov spaces}



\maketitle

\section{Introduction}\label{sec1}

%

Gravitational waves are currently both an object of academic interest and an object of experimental observation, which provides new data about the Universe
\cite{PhysRevLett.116.061102,PhysRevX.9.031040,PhysRevX.11.021053}.
In addition, from the point of view of theoretical cosmology, the discovery of primordial gravitational waves can provide a weighty argument in support of the inflationary model of the initial stage of the Universe.
Obtaining exactly integrable models in this direction, as the basic elements of the theory, makes a significant contribution to the development of gravitational-wave research.

This paper presents new results on primordial gravitational waves in Bianchi type VII spaces, which are spatially homogeneous but not isotropic like the standard Friedmann–Robertson–Walker model, inspired by observational data on the anisotropy of the electromagnetic microwave background of the Universe \cite{Bennett2013} and is of interest for studying the initial stages of the dynamics of the Universe.

The main manifestation of the gravitational field from the point of view of the motion of test particles is the deviation of the geodesics along which the particles move. The deviation of geodesics leads to the appearance of tidal accelerations in the gravitational wave, which can cause various physical effects. For example, the primordial gravitational wave, acting on the primary plasma, excites plasma density waves, which leads to the appearance of additional electromagnetic fields that contributed to the characteristics of the background microwave radiation, which is currently detected using satellite telescopes \cite{Bennett2013}.

The calculation in such problems is quite complicated and is carried out mainly by approximate and numerical methods. In this regard, finding exact solutions to the equations of deviation of geodesics in a gravitational wave seems to be an important basic element of the theory, which makes it possible to exact calculate tidal accelerations and the resulting physical effects. Exact solutions in any physical theory are the cornerstones that make it possible to understand the qualitative behavior of physical objects in more complex cases. In addition, exact solutions make it possible to effectively debug approximate and numerical methods for more complex physical models, as well as to train computer ''artificial intelligence'' in recognizing gravitational waves.

As an initial model for a gravitational wave, we use Shapovalov type III wave spaces \cite{Osetrin2020Symmetry} in this paper, which allow the existence of privileged coordinate systems, where the metric depends on only one wave variable, along which the space-time interval vanishes. Shapovalov's wave spaces make it possible to construct a complete integral of the equation of motion of test particles in the Hamilton-Jacobi formalism and, consequently, to find the trajectories of motion of test particles in a gravitational wave in quadratures. The presence of the complete integral of the Hamilton-Jacobi equation also allows one to obtain the exact solution of the geodesic deviation equations, and then calculate the tidal accelerations for particles in a gravitational wave.

The privileged coordinate systems used to integrate the Hamilton-Jacobi equation are based on the three commuting Killing vectors allowed by the Shapovalov type III wave space, which determine the so-called "complete set" of integrals of the motion of test particles and make it possible to carry out analytical integration of the Hamilton-Jacobi equation by the method of complete separation of variables \cite{ Shapovalov1978I, Shapovalov1978II, Shapovalov1979}.

An additional advantage of Shapovalov wave spaces is the ability to construct an explicit coordinate transformation from a privileged coordinate system, where the exact integration of the equations of motion of test particles in the Hamilton-Jacobi formalism is allowed, into synchronous reference systems, where time and spatial coordinates are separated, and an observer freely falling in a gravitational field is at rest . Synchronous reference systems make it possible to synchronize time at different points of space-time and clarify the physical content of the obtained exact models.

Spaces that allow the existence of privileged coordinate systems, where the Hamilton-Jacobi equation of test particles can be integrated by the method of complete separation of variables, are widely used in the theory of gravity, since they allow integrating field equations and obtaining exactly integrable models. These are models with an electromagnetic field \cite{Obukhov1988,Obukhov2022632,Obukhov2021695},
models with dust and pure radiation \cite{OsetrinDust2016,OsetrinVaidya1996,OsetrinVaidya2009,OsetrinRadiation2017},
models with a scalar field \cite{OsetrinScalar2018,Obukhov2022142,Obukhov2021134,Obukhov2021183}.
Obtaining realistic models of primordial gravitational waves of the Universe on the basis of Shapovalov's wave spaces is also possible for gravity models taking into account quantum and other corrections (for example, f(R) - gravity, etc. \cite{Odintsov2007,Odintsov2011,Capozziello2011,Odintsov2017}).

Shapovalov's wave spaces make it possible to obtain exact wave solutions in the theory of gravity not only in Einstein's general theory of relativity, but also in modified theories of gravity \cite{OsetrinScalar312020,OsetrinScalar212020,OsetrinSymmetry2021}, which makes it possible to compare analogous solutions and select realistic theories and models.

Shapovalov's wave spaces allow selection of classes of spatially homogeneous models, which are interpreted as cosmological models of the Universe and, accordingly, describe primordial gravitational waves. So, for example, type III Shapovalov spaces allow IV, VI and VII types of Bianchi Universes (see for example \cite{OsetrinHomog312002,OsetrinHomog2006}), type II Shapovalov spaces allow Bianchi universes of type III \cite{OsetrinHomog212020}.

In this paper, we will consider the Shapovalov wave spaces related to Bianchi Universes of type VII.
Trajectories of test particles will be found, exact solutions of the geodesic deviation equations in a gravitational wave will be found, and tidal accelerations of particles in a gravitational wave will be calculated. The results are presented both in a privileged coordinate system, where the equations for test particle trajectories are integrated, and in a laboratory synchronous coordinate system associated with a freely falling observer.

\section{
Shapovalov wave spacetimes
}

The Hamilton-Jacobi equation for a test particle of mass $m$ in a gravitational field with the metric tensor $g_{{\alpha}{\beta}}(x^{\gamma})$
has the form (see \cite{LandauEng1}):
\begin{equation}
g^{{\alpha}{\beta}}\frac{\partial S}{\partial x^{\alpha}}\frac{\partial S}{\partial x^{\beta}}=m^2c ^2
,
\label{HJE}
\end{equation}
$$
{\alpha},{\beta},{\gamma} ...
=0,...,(n-1),
$$
where the capital letter $S$ denotes the test particle action function, $n$ is the space dimension. We will set the speed of light $c$ to be equal to unity.
In what follows, we will also use the lowercase letter $s$ to denote a space-time interval.
\begin{definition}
\label{stack_space}
If the space allows the existence of a privileged coordinate system $\{x^{\alpha}\}$, where the Hamilton-Jacobi equation (\ref{HJE}) can be integrated by the complete separation of variables method, when the complete integral $S$ for the test particle action function can be written in ''separated'' form:
\begin{equation}
S=\sum_{\alpha=0}^{n-1}\phi_\alpha(x^\alpha,\lambda_0,...,\lambda_{n-1})
,
\end{equation}
$$
\lambda_0,...,\lambda_{n-1}-\mbox{const},
\qquad
\det \Bigl\lvert \frac{\partial{}^2 S}{\partial x^{\alpha}\partial \lambda_{\beta}}\Bigr\rvert \ne 0
,
$$
moreover, one of the non-ignored variables (on which the metric depends) is a wave (null), i.e. the space-time interval along this variable vanishes, then such a space is called the Shapovalov wave space.
\end{definition}

Shapovalov spaces represent gravitational-wave models of space-time, which allow integrating the equations of motion of test particles and radiation in the Hamilton-Jacobi formalism in quadratures, i.e. make it possible to find an explicit form of geodesic lines along which test particles move.

Shapovalov wave spaces are a subclass of wave models for the broader class of {St{\"{a}}ckel} spaces (Paul {St{\"{a}}ckel}, see \cite{Stackel1891thesis,Stackel1897145}), which allow separation of variables in the Hamilton-Jacobi equation. The theory of {St{\"{a}}ckel} spaces was built by the efforts of a large number of researchers and was basically completed by the construction of a classification and a detailed description of these spaces in the works of Vladimir Shapovalov \cite{Shapovalov1978I,Shapovalov1978II,Shapovalov1979}.

Shapovalov spaces admit a ''complete set'' of Killing vectors and Killing tensors of the second rank, which correspond to sets of integrals of motion of test particles. Shapovalov wave spaces are classified according to the Abelian group of motions they admit, which specifies the number of ignored variables (on which the metric in the privileged coordinate system does not depend) and, in the case of space-time, can admit from one to three commuting Killing vectors in a ''complete set'' and, accordingly, for the case of space-time there are types I, II and III of Shapovalov wave spaces.

Shapovalov's wave spaces allow explicit construction of the complete integral of the Hamilton-Jacobi equation for test particles and make it possible to obtain an analytical form of their motion trajectories in the Hamilton-Jacobi formalism.
The test particle coordinates $x^{\alpha}$ are functions of proper time $\tau$ on the base geodesic line along which the test particle moves, and is given in the Hamilton-Jacobi formalism by the particle trajectory equations in the form:
\begin{equation}
\label{MovEqu}
\frac{\partial S (x^{\alpha},\lambda_{k})}{\partial \lambda_{j}}=\sigma_{j},
\quad
\tau=S (x^{\alpha},\lambda_{k})\rvert_{m=1},
\end{equation}
$$
{i},{j},{k}=1,2,3;
$$
where $\lambda_{k}$, $\sigma_{k}$ are independent constant parameters determined by the initial or boundary data for the motion of a test particle along the base geodesic line, the variable $\tau$ is the proper time of the particle.

Shapovalov type III wave space-time metric in a privileged coordinate system depends on one wave variable and can be reduced to the following form
(see f.e. \cite{OsetrinHomog312002,OsetrinHomog2006,OsetrinHomog312020}):
$$
{ds}^2=2dx^0dx^1
+g_{ab}(x^0)\Bigl(dx^a+g^a(x^0) \, dx^1\Bigr)
$$
\begin{equation}
\Bigl(dx^b +g^b(x^0) \, dx^1\Bigr)
\label{MetricShapovalovIII}
,\end{equation}
where indices a, b run through the values 2, 3. Thus, in the general case, the metric includes five arbitrary functions of the wave (null) variable $x^0$.

Einstein's equations with cosmological constant $\Lambda$ in vacuum
\begin{equation}
R_{\alpha\beta}=\Lambda g_{\alpha\beta}
\label{EinsteinEqs}
\end{equation}
for the metric (\ref{MetricShapovalovIII}) lead to the following necessary conditions:
\begin{equation}
\Lambda=g^a=0
.\end{equation}
The space-time metric for a gravitational wave (\ref{MetricShapovalovIII}) has type N according to Petrov's classification.

\section{
Deviation of geodesics in Shapovalov wave spaces
}

The deviation of geodesics in the general theory of relativity and in modified theories of gravity is the main manifestation of the gravitational field and gravitational waves in particular. Geodesic deviation is used to detect the gravitational field and provides information about the space-time curvature tensor, and also allows you to calculate various effects of gravity on physical objects, incl. tidal effects of the gravitational field.

The geodesic deviation equation has the following form:
\begin{equation}
\label{deviation}
\frac{D^2 \eta^{\alpha}}{{d\tau}^2}=
 R^{\alpha}{}_{{\beta}{\gamma}{\delta}}u^{\beta} u^{\gamma}\eta^{\delta}.
\end{equation}
Here
$u^{\alpha}(\tau)$ is the four-velocity of the test particle on the base geodesic line,
$D$ is the covariant derivative,
$\eta^{\alpha}(\tau)$ is the geodesic deviation vector, $R^{\alpha}{}_{{\beta}{\gamma}{\delta}}$ is the Riemann curvature tensor.
This equation is considered along the base geodesic line of a test particle with four-velocity $u^{\alpha}$ and, therefore, the coordinates $x^{\alpha}$ are parametrized by one parameter (the proper time of the test particle $\tau$).

For the four-velocity test particle, the normalization condition is satisfied
\begin{equation}
\label{Norm}
g^{{\alpha}{\beta}}u_{\alpha} u_{\beta} =1.
\end{equation}
For four-velocity, assuming the mass of the test particle to be equal to unity, we obtain the expression:
\begin{equation}
\label{General4U}
u_{\alpha}=\frac{\partial S}{\partial x^{\alpha}}
\,\biggl\rvert_{m=1}
,\end{equation}
where $S$ is the test particle action function.

Thus, first finding the form of the metric from the field equations for the corresponding gravitational model, then obtaining the trajectories of the test particles from the Hamilton-Jacobi equations (\ref{HJE}) and calculating the four-velocities of the test particles from the relation (\ref{General4U}), one can write the deviation equations (\ref{deviation}) explicitly and integrate the resulting system of differential equations in a direct way.

On the other hand, if we are able to find the complete integral of the Hamilton-Jacobi equation for a test particle
$S (x^{\alpha} ,\lambda_{a})$,
then the deviation vector of geodesics
$\eta^{\alpha}$
can be found also
according to the method proposed in the work
\cite{Bazanski19891018}. Using the variational method, a generalization of the Hamilton-Jacobi equations for the complete integral of geodesics was found, which made it possible to obtain equations for determining the deviation of geodesics through the complete integral of the Hamilton-Jacobi equation.
According to this method, the deviation vector $\eta^{\alpha}$ can be found as a solution of a linear algebraic system of equations on the trajectory of the ''base'' test particle
\begin{equation}
\label{Deviation1}
\eta^{\alpha}\,
\frac{\partial u_{\alpha}(x^{\beta},\lambda_{j})}{\partial\lambda_{i}} 
+\rho_{k}\frac{\partial^2 S(x^{\beta},\lambda_{j})}{\partial\lambda_{i}\partial\lambda_{k}}=\vartheta_{i}
,
\end{equation} 
\begin{equation}
\label{Deviation2}
u_{\alpha}(x^{\beta},\lambda_{i})\,\eta^{\alpha}=0
,
\end{equation}
$$
{i},{j},{k} = 1\ldots 3;
\qquad
{\alpha}, {\beta}, {\gamma}=0\ldots 3,
$$
where $\lambda_{i}$, $\rho_{i}$, $\vartheta_{i}$ are independent constant parameters.
Here the constants $\lambda_{i}$ are determined by the initial data or boundary conditions for the velocity (momentum) of the test particle on the base geodesic line.
The constants $\rho_{i}$ are determined by the initial data for the deviation vector (that is, they are related to the initial data adjacent geodetic line). 
The values ot the constants $\vartheta_{i}$ are related to the definition of the origin of the variables $x^{\alpha}$. 
The equation (\ref{Deviation2}) expresses the condition that the deviation vector $\eta^{\alpha}$ is orthogonal to the base geodesic line.The dependence of the variables $x^{\alpha}$ on the proper time $\tau$ on the base geodesic line along which the test particle moves is established in the Hamilton-Jacobi formalism by the equations (\ref{MovEqu}) for the particle trajectory.

Thus, the Shapovalov wave spaces make it possible to obtain exact solutions for gravitational waves both of the equations of motion of test particles and to find exact solutions of the geodesic deviation equations in a gravitational wave.

\section{Spatially homogeneous models of Shapovalov spaces}

Earlier, in the study of spatially homogeneous symmetries of Shapovalov spaces of type III
 it was shown that three types of spatially homogeneous models can be built on the basis of these spaces: types IV, VI and VII according to Bianchi (see f.e. \cite{OsetrinHomog312002,OsetrinHomog2006,OsetrinHomog312020}). Moreover, gravitational waves in Shapovalov spaces of type IV and VI types of Bianchi are aperiodic in the wave variable.

Bianchi type VII Shapovalov wave space metric
in a privileged coordinate system, where the metric depends only on the wave (null) variable $x^0$ (along the variable $x^0$ the space-time interval vanishes), can be represented as \cite{OsetrinHomog2006}:
$$
{ds}^2=2dx^0dx^1
$$
$$
\mbox{}
-
\frac{{x^0}^{2 \omega } }{\gamma  \left({\delta}^2-1\right)}
\,\biggl[
\left(1-{\delta} \cos \left(\theta -2\log {x^0}\right)\right) \,\left({dx^2}\right)^2
$$
$$
\mbox{}
-
2  \sin \left(\theta -2\log {x^0}\right)\,{dx^2}{dx^3}
$$
\begin{equation}
\mbox{}
+
\left(1+{\delta} \cos \left(\theta -2\log {x^0}\right)\right)
 \,\left({dx^3}\right)^2
\,\biggr]
\label{Metric}
,\end{equation}
\begin{equation}
{g=\det g_{ij}=}\frac{-{x^0}^{4 \omega }}{\gamma ^2 \left({1-\delta}^2\right)}
,\qquad
{\delta}^2<1
,\end{equation}
where $x^0$ is the wave variable,
the constants ${\gamma}$, ${\delta}$, $\theta$ and ${\omega}$ 
are independent parameters of the gravitational wave model. The parameter ${\omega}$ is the parameter of the Bianchi type VII spatial homogeneity group.

The space-time model under consideration can be interpreted as a model of a propagating primordial gravitational wave against the background of an expanding Universe.
 
The space with the metric (\ref{Metric}) admits a covariantly constant vector $K$ and, therefore, is a plane-wave space:
\begin{equation}
\nabla_{\beta} K_{\alpha}=0
\quad
\to
\quad
K_{\alpha}=\bigl( K_0,0,0,0 \bigl)
,\end{equation}
where $K_0$ is a {constant}.

A space with a metric (\ref{Metric}) admits a spatial homogeneity group with Killing vectors $X_{(1)}$, $X_{(2)}$, and $X_{(3)}$, which can be chosen in the privileged coordinate system in the form
\begin{equation}
X^{\alpha}_{(1)}=\bigl(0,0,1,0\bigr),
\qquad
X^{\alpha}_{(2)}=\bigl(0,0,0,1\bigr),
\end{equation}
\begin{equation}
X^{\alpha}_{(3)}=\bigl(-x^0, \,x^1, \,\omega\, x^2-x^3, \,x^2+\omega\,x^3 \bigr)
.\end{equation}
The additional fourth Killing vector associated with the choice of the privileged coordinate system commutes with the vectors $X_{(1)}$ and $X_{(2)}$ and has the form in this coordinate system
\begin{equation}
X^{\alpha}_{(0)}=\bigl(0,1,0,0\bigr)
.
\end{equation}
The Killing vector $X_{(0)}$ is a null vector, because $g_{{\alpha}{\beta}}X^{\alpha}_{(0)}X^{\beta}_{(0)}=0$.

The commutation relations for the Killing vectors have the form:
\begin{equation}
\left[X_{(0)},X_{(1)}\right]=0
,\qquad
\left[X_{(0)},X_{(2)}\right]=0
,\end{equation}
\begin{equation}
\left[X_{(0)},X_{(3)}\right]=X_{(0)}
,\end{equation}
\begin{equation}
\left[X_{(1)},X_{(2)}\right]=0
,\end{equation}
\begin{equation}
\left[X_{(1)},X_{(3)}\right]= {\omega}X_{(1)}+X_{(2)}
,\end{equation}
\begin{equation}
\left[X_{(2)},X_{(3)}\right]=-X_{(1)}+ {\omega}X_{(2)}
.
\end{equation}
The Killing vectors $X_{(0)}$, $X_{(1)}$, $X_{(2)}$ and $X_{(3)}$ generate a 4-dimensional motion group, the $X_{(0 )}$, $X_{(1)}$, $X_{(2)}$ generate a 3-dimensional abelian subgroup, vectors $X_{(1)}$, $X_{(2)}$ and $X_{ (3)}$ generate a 3-dimensional subgroup of the spatial homogeneity of the type VII Bianchi model.

\section{Gravitational Wave for Type VII Bianchi Models in Einstein's Theory of Gravity}

Consider the solution of the Einstein equations in vacuum (\ref{EinsteinEqs}) for the gravitational wave metric (\ref{Metric}).
As a result, we obtain the following exhaustive restrictions on the parameters of the gravitational wave:
\begin{equation}
\label{deltaParameterRange}
{\delta}^2 = \frac{\omega (\omega -1) }{\omega ^2-\omega -1}
,\end{equation}
\begin{equation}
\label{omegaParameterRange}
0\le\omega\le 1
,\qquad
0\le\delta^2\le 1/5
.\end{equation}
Thus, three independent constant parameters of the gravitational wave remain in the model under consideration: the Bianchi type VII homogeneity subgroup parameter $\omega$, the $\gamma$ parameter related to the wave amplitude, and the angular parameter $\theta$ related to the wave phase.

The $(+,-,-,-)$ metric signature imposes additional restrictions on the choice of $\gamma$ sign:
\begin{equation}
\gamma<0.
\end{equation}

To shorten the formulas, we will further use the notation ${\delta}$, understanding it as an expression of the following form:
\begin{equation}
\label{deltaOmega}
{\delta}(\omega) =\pm \sqrt{ \frac{\omega (\omega -1) }{\omega ^2-\omega -1} }
.\end{equation}

The nonzero components of the Riemann curvature tensor $R_{\alpha\beta\gamma\delta}$ for the metric (\ref{Metric}), subject to the constraints (\ref{deltaParameterRange}) and (\ref{omegaParameterRange}) in the privileged coordinate system, take the following form:
$$
{R}_{0202} = 
-\frac{\left(\omega ^2-\omega -1\right) {x^0}^{2 \omega -2} 
}{\gamma }
\Bigl[
-2 (\omega -1) \omega 
$$
$$
\mbox{}
+
2 \left(\omega ^2-\omega -1\right) {\delta} \cos \left( \theta -2\log {x^0} \right)
$$
\begin{equation}
\mbox{}
+(2 \omega -1) {\delta} \sin \left( \theta -2\log {x^0} \right)
\Bigr]
,\end{equation}
$$
{R}_{0302} = 
\frac{\left(\omega ^2-\omega -1\right) {\delta} {x^0}^{2 \omega -2} 
}{\gamma }
\Bigl[
$$
$$
(2 \omega -1) \cos \left( \theta -2\log {x^0} \right)
$$
\begin{equation}
\mbox{}
+
2 \left(-\omega ^2+\omega +1\right) \sin \left( \theta -2\log {x^0} \right)
\Bigr]
,\end{equation}
$$
{R}_{0303} = \frac{\left(\omega ^2-\omega -1\right) {x^0}^{2 \omega -2} 
}{\gamma }
\Bigl[
2 (\omega -1) \omega 
$$
$$
\mbox{}
+
2 \left(\omega ^2-\omega -1\right) {\delta} \cos \left( \theta -2\log {x^0} \right)
$$
\begin{equation}
\mbox{}
+(2 \omega -1) {\delta} \sin \left( \theta -2\log {x^0} \right)
\Bigr]
.\end{equation}
Thus, for the value of the parameter $\omega=0,1$ the parameter $\delta$ and the Riemann curvature tensor $R_{\alpha\beta\gamma\delta}$
vanish, the model degenerates, and space-time becomes flat.

\section{Integration of the Hamilton-Jacobi equation for test particles (case of $\omega\ne 1/2$)}

To obtain the trajectories of motion of test particles in a gravitational wave (\ref{Metric}), we consider the Hamilton-Jacobi equation (\ref{HJE}) in the given space. In accordance with the general properties of Shapovalov spaces, we will look for the complete integral of the Hamilton-Jacobi equation in a separated form:
\begin{equation}
S(x^\alpha,\lambda_k)=\phi_0(x^0)+\sum_{k=1}^3 \lambda_k x^k
,\end{equation}
where the independent constant parameters $\lambda_k$ are determined by the initial or boundary conditions for the motion of a test particle.

Then the Hamilton-Jacobi equation (\ref{HJE}) gives for the function $\phi_0(x^0)$ (${\lambda_1}\ne0$):
$$
 \phi_0{}' = 
 \frac{1}{2 {\lambda_1}} 
 +
 \frac{\gamma {x^0}^{-2 \omega } 
 }{2 {\lambda_1}} 
\biggl[
 -{\lambda_2}^2   -{\lambda_3}^2 
$$
$$
\mbox{}
  -2{\delta}  {\lambda_2} {\lambda_3}  \sin \left(\theta -2\log {x^0}\right)
 $$
\begin{equation}
\label{PhiEq}
\mbox{}
- {\delta} \left({\lambda_2}^2-{\lambda_3}^2\right) \cos \left(\theta -2\log {x^0}\right) 
\biggr]
.\end{equation}

The integration of the equation (\ref{PhiEq}) has a singularity at $\omega=1/2$. Therefore, we will further consider the case $\omega=1/2$ separately.

In this section, we will assume that $\omega\ne 1/2$. Then the equation (\ref{PhiEq}) gives 
for the function $\phi_0(x^0)$:
$$
 \phi_0 (x^0) = 
 \frac{x^0}{2 {\lambda_1} }
 +
 \frac{
 {x^0}^{1-2 \omega } \gamma({\lambda_2}^2 +{\lambda_3}^2 )
  }{2 {\lambda_1}  (2 \omega -1) } 
 $$
 $$
 \mbox{}
 +
 \frac{
 {\gamma} {\delta} 
 {x^0}^{1-2 \omega } 
  }{2 {\lambda_1}  
  \left(4 \omega ^2-4 \omega +5\right)} 
\Bigl[
$$
$$
 2 
  \left({\lambda_2}^2
 +{\lambda_2} {\lambda_3} 
(2 \omega -1)
 -{\lambda_3}^2\right) \sin \left(\theta -2\log {x^0}\right)
 $$
\begin{equation}
\mbox{}
 +
\Bigl(
 (2 \omega -1)
 \left( {\lambda_2}^2+{\lambda_3}^2 \right) 
 -4 {\lambda_2} {\lambda_3}
 \Bigr)
 \cos \left(\theta -2\log {x^0}\right)
\Bigr]
. \end{equation}
Thus, we have found the complete integral of the Hamilton-Jacobi equation of test particles $S(x^\alpha,\lambda_k)$.

Now we can write and solve particle trajectory equations (\ref{MovEqu}) in the Hamilton-Jacobi formalism.

\onecolumn

Omitting obvious calculations, we present the result of solving the equations (\ref{MovEqu}) in the form of a test particle trajectory in a privileged coordinate system:
\begin{equation}
x^0 (\tau)  = {\lambda_1} {\tau} 
, \end{equation}
$$
x^1 (\tau)  = 
\frac{
({\lambda_1} {\tau})^{1-2 \omega } 
}{2 {\lambda_1}^2
(2 \omega -1 )
 } 
\Bigl(
(2 \omega -1) ({\lambda_1} {\tau})^{2 \omega }
+\gamma({\lambda_2}^2  +{\lambda_3}^2 )
\Bigr)
$$
$$
\mbox{}
+
\frac{
\gamma  {\delta} 
({\lambda_1} {\tau})^{1-2 \omega} 
}{2 {\lambda_1}^2
 (4 \omega ^2 - 4 \omega + 5)} 
\Bigl[
2 
\Bigl(
{\lambda_2}^2+{\lambda_2} {\lambda_3} (2 \omega -1)-{\lambda_3}^2
 \Bigr)
\sin \left(\theta -2\log \left({\lambda_1} {\tau}\right)\right)
$$
\begin{equation}
\mbox{}
+
\Bigl(
{\lambda_2}^2 (2 \omega -1)-4 {\lambda_2} {\lambda_3}+{\lambda_3}^2 (1-2 \omega )
\Bigr)
\cos \left(\theta -2\log \left({\lambda_1} {\tau}\right)\right)
\Bigr]
, \end{equation}
$$
x^2 (\tau)  = 
-\frac{
\gamma  
 ({\lambda_1} {\tau})^{1-2 \omega } 
}{{\lambda_1} (2 \omega -1 ) (4 \omega ^2 - 4 \omega + 5)}
\Bigl[
{\lambda_2} \left(4 \omega ^2-4 \omega +5\right)
$$
$$
\mbox{}
+
(2 \omega -1) {\delta} (2 {\lambda_2}+{\lambda_3} (2 \omega -1)) \sin \left(\theta -2\log \left({\lambda_1} {\tau}\right)\right)
$$
\begin{equation}
+(2 \omega -1) {\delta} ({\lambda_2} (2 \omega -1)-2 {\lambda_3}) \cos \left(\theta -2\log \left({\lambda_1} {\tau}\right)\right)
\Bigr]
, \end{equation}
$$
x^3 (\tau)  = 
\frac{\gamma  
({\lambda_1} {\tau})^{1-2 \omega } 
}{{\lambda_1}(2 \omega -1 ) (4 \omega ^2 - 4 \omega + 5)} 
\Bigl[
{\lambda_3} \left(-4 \omega ^2+4 \omega -5\right)
$$
$$
\mbox{}
-(2 \omega -1) {\delta} ({\lambda_2} (2 \omega -1)-2 {\lambda_3}) \sin \left(\theta -2\log \left({\lambda_1} {\tau}\right)\right)
$$
\begin{equation}
\mbox{}
+(2 \omega -1) {\delta} (2 {\lambda_2}+{\lambda_3} (2 \omega -1)) \cos \left(\theta -2\log \left({\lambda_1} {\tau}\right)\right)
\Bigr]
, \end{equation}
where $\tau$ is the proper time of the test particle. In the process of integrating the equations (\ref{MovEqu}), we set the constants $\sigma_k$ equal to zero by choosing the origin of the variables $x^\alpha$ and the proper time $\tau$.

Obtaining an explicit form of test particle trajectories in a gravitational wave here demonstrates the possibilities of exact integration of wave models in privileged coordinate systems in Shapovalov spaces.

For the solutions obtained, we find the 4-velocity of the test particle in the privileged coordinate system:
\begin{equation}
u^{\alpha}(\tau) =\frac{D x^{\alpha}}{d\tau}
, \end{equation}
\begin{equation}
u^{0} = {\lambda_1} 
,\qquad
x^0= {\lambda_1} \tau
\label{uPrivilUp0}
, \end{equation}
\begin{equation}
u^{1} = 
\frac{
{x^0}^{-2 \omega } 
 }{2 {\lambda_1}} 
\Bigl[
{x^0}^{2 \omega }
-\gamma  {\delta} \left({\lambda_2}^2-{\lambda_3}^2\right) \cos \left(\theta -2\log {x^0}\right)
-2 {\lambda_2} {\lambda_3} \gamma  {\delta} \sin \left(\theta -2\log {x^0}\right)
-\gamma ({\lambda_2}^2 +{\lambda_3}^2)
\Bigr]
\label{uPrivilUp1}
, \end{equation}
\begin{equation}
\label{uPrivilUp2}
u^{2} = \gamma  {x^0}^{-2 \omega } 
\left[
{\lambda_2} {\delta} \cos \left(\theta -2\log {x^0}\right)
+{\lambda_3} {\delta} \sin \left(\theta -2\log {x^0}\right)
+{\lambda_2}
\right]
, \end{equation}
\begin{equation}
\label{uPrivilUp3}
u^{3} = \gamma  {x^0}^{-2 \omega } 
\left[
{\lambda_2} {\delta} \sin \left(\theta -2\log {x^0}\right)
-{\lambda_3}{\delta} \cos \left(\theta -2\log {x^0}\right)+{\lambda_3}
\right]
, \end{equation}

Having obtained complete the integral of the Hamilton-Jacobi equation $S(x^\alpha,\lambda_k)$ and the 4-velocity of test particles, we can now write and solve equations for the components of the geodesic deviation vector
(\ref{Deviation1})-(\ref{Deviation2}).

Omitting the obvious calculations for solving the system of algebraic equations (\ref{Deviation1})-(\ref{Deviation2}), we present the exact solution for the geodesic deviation vector $\eta^\alpha (\tau)$ in the privileged coordinate system ($\omega\ne 1/2$):
\begin{equation}
\label{DeviationSolutionPriv0}
 \eta^0 (\tau)  = {\rho_1} \tau -{\lambda_1} {\Omega} 
 ,\qquad
x^0(\tau)=\lambda_1\tau
, \end{equation}
$$
 \eta^1 (\tau)  = 
 {\vartheta_1}
 -\frac{
\gamma  {R_2} {x^0}^{1-2 \omega } 
}{{\lambda_1}^3 (2 \omega -1) \left(4 \omega ^2-4 \omega +5\right)}
\Bigl[
{\lambda_2} \left(4 \omega ^2-4 \omega +5\right)
$$
$$
+
(2 \omega -1) {\delta} (2 {\lambda_2}+{\lambda_3} (2 \omega -1)) \sin \left(\theta -2\log {x^0}\right)
$$
$$
+(2 \omega -1) {\delta} ({\lambda_2} (2 \omega -1)-2 {\lambda_3}) \cos \left(\theta -2\log {x^0}\right)
\Bigr]
$$
$$
 +
 \frac{
 \gamma  {R_3}{x^0}^{1-2 \omega } 
  }{{\lambda_1}^3 (2 \omega -1) \left(4 \omega ^2-4 \omega +5\right)}
\Bigl[
{\lambda_3} \left(-4 \omega ^2+4 \omega -5\right)
$$
$$
 -(2 \omega -1) {\delta} ({\lambda_2} (2 \omega -1)-2 {\lambda_3}) \sin \left(\theta -2\log {x^0}\right)
$$
$$
 +(2 \omega -1) {\delta} (2 {\lambda_2}+{\lambda_3} (2 \omega -1)) \cos \left(\theta -2\log {x^0}\right)
\Bigr]
 $$
 $$
 +\frac{
 {x^0}^{-2 \omega } 
  }{2 {\lambda_1}^3} 
\Bigl[
\gamma  {\delta} \left({\lambda_2}^2-{\lambda_3}^2\right) \left({\lambda_1}^2 {\Omega}-{\rho_1} {x^0}\right) \cos \left(\theta -2\log {x^0}\right)
 $$
 $$
 +2 {\lambda_2} {\lambda_3} \gamma  {\delta} \left({\lambda_1}^2 {\Omega}-{\rho_1} {x^0}\right) \sin \left(\theta -2\log {x^0}\right)
  -{\rho_1} {x^0}^{2 \omega +1}
 $$
\begin{equation}
\mbox{}
- {\lambda_1}^2{\Omega} {x^0}^{2 \omega }
+({\lambda_2}^2+{\lambda_3}^2 )\gamma ({\lambda_1}^2{\Omega}- {\rho_1} {x^0})
\Bigr]
\label{DeviationSolutionPriv1}
, \end{equation}
$$
\eta^2 (\tau)  =
{\vartheta_2} 
-\frac{
\gamma  {x^0}^{-2 \omega } 
\left(
{\lambda_1}^2 {\Omega}-{\rho_1} {x^0}
\right) 
}{{\lambda_1}^2}
\Bigl[
{\lambda_2}
+{\lambda_2} {\delta} \cos \left(\theta -2\log {x^0}\right)
$$
$$
+{\lambda_3} {\delta} \sin \left(\theta -2\log {x^0}\right)
\Bigr]
$$
$$
+\frac{
\gamma  {R_2} {x^0}^{1-2 \omega } 
}{{\lambda_1}^2 (2 \omega -1) \left(4 \omega ^2-4 \omega +5\right)}
\Bigl[
4 \omega ^2-4 \omega +5
$$
$$
+2 (2 \omega -1) {\delta} \sin \left(\theta -2\log {x^0}\right)
+(1-2 \omega )^2 {\delta} \cos \left(\theta -2\log {x^0}\right)
\Bigr]
$$
\begin{equation}
-\frac{\gamma  {R_3} {\delta} {x^0}^{1-2 \omega } 
\left[
(1-2 \omega ) \sin \left(\theta -2\log {x^0}\right)+2 \cos \left(\theta -2\log {x^0}\right)
\right]
}{{\lambda_1}^2 \left(4 \omega ^2-4 \omega +5\right)}
\label{DeviationSolutionPriv2}
, \end{equation}
$$
 \eta^3 (\tau)  = 
 {\vartheta_3}
 +
 \frac{\gamma  {x^0}^{-2 \omega } \left({\lambda_1}^2 {\Omega}-{\rho_1} {x^0}\right) 
  }{{\lambda_1}^2}
\Bigl[
 -{\lambda_2} {\delta} \sin \left(\theta -2\log {x^0}\right)
 $$
 $$
 +{\lambda_3} {\delta} \cos \left(\theta -2\log {x^0}\right)-{\lambda_3}
\Bigr]
 $$
 $$
 -\frac{\gamma  {R_2} {\delta} {x^0}^{1-2 \omega } 
 \left[
 (1-2 \omega ) \sin \left(\theta -2\log {x^0}\right)+2 \cos \left(\theta -2\log {x^0}\right)
 \right]
 }{{\lambda_1}^2 \left(4 \omega ^2-4 \omega +5\right)}
 $$
 $$
 -\frac{
 \gamma  {R_3} {x^0}^{1-2 \omega } 
  }{{\lambda_1}^2 (2 \omega -1) \left(4 \omega ^2-4 \omega +5\right)}
\Bigl[
 -4 \omega ^2+4 \omega -5
 $$
\begin{equation}
 +2 (2 \omega -1) {\delta} \sin \left(\theta -2\log {x^0}\right)
 +(1-2 \omega )^2 {\delta} \cos \left(\theta -2\log {x^0}\right)
\Bigr]
\label{DeviationSolutionPriv3}
, \end{equation}
where, for brevity, the auxiliary notation $R_2$, $R_3$ and $\Omega$ is introduced:
\begin{equation}
\label{R23}
R_2 = {\lambda_2} {\rho_1}-{\lambda_1} {\rho_2}
,\qquad
R_3 = {\lambda_3} {\rho_1}-{\lambda_1} {\rho_3}
,\end{equation}
\begin{equation}
\label{constOmega}
\Omega = {\lambda_1} {\vartheta_1}+{\lambda_2} {\vartheta_2}+{\lambda_3} {\vartheta_3}
.\end{equation}

The exact solution obtained above for the deviation vector $\eta^\alpha (\tau)$ includes a number of independent parameters.
The parameters $\omega$, $\gamma$ and $\theta$ determine the characteristics of the gravitational wave.
The parameters $\lambda_k$ define the motion of the test particle on the base geodesic and are determined by the initial or boundary conditions for the particle velocity on the base geodesic.
The parameters ${\rho_k}$ and ${\vartheta_k}$ are determined by the initial or boundary conditions for the velocity and relative position of the particle on the adjacent geodesic.


One of the important manifestations of geodesic deviation in a gravitational field is tidal accelerations $A^\alpha = D^2\eta^\alpha/d^2\tau$, which affect physical objects in a gravitational wave.

In a privileged coordinate system, the form of tidal accelerations in a gravitational wave is quite cumbersome, so we will not present it here, we only note that
component $A^0=0$. Further, passing to the synchronous frame of reference, we will find a compact form for tidal accelerations.

\section{Gravitational wave in synchronous frame of reference (case of~$\omega\ne 1/2$)}

Synchronous frames of reference are physically distinguished frames of reference, since they allow one to synchronize time at different points in space-time \cite{LandauEng1}.
Also, in contrast to the privileged coordinate system, which we used to integrate the Hamilton-Jacobi equations, in the synchronous reference frame, time and spatial coordinates are separated, which is important for the observer. Therefore, the possibility of passing to a synchronous frame of reference is an advantage in the physical analysis of solutions. Shapovalov wave spaces provide such an opportunity, since the presence of the complete integral of the Hamilton-Jacobi equation and the knowledge of trajectories for test particles allows one to find an explicit analytical form of such a transition. Moreover, the proper time of an observer freely falling along the base geodesic $\tau$ in a synchronous frame of reference becomes a common time variable. According to the algorithm described in \cite{LandauEng1}, the transformation from the privileged coordinate system $x^\alpha$ to the synchronous frame $\tilde x{}^\alpha$ has the form:
\begin{equation}
x^{\alpha} \to \tilde x{}^{\alpha}=\left(\tau,\lambda_1,\lambda_2,\lambda_3 \right)
, \end{equation}
\begin{equation}
x^0 = {\tilde x{}^1} {\tau} 
\label{ToSynchr0}
, \end{equation}
$$
x^1 = 
\frac{
\tau^2
({\tilde x{}^1} {\tau})^{-2 \omega-1 } 
}{2 
(2 \omega -1 ) (4 \omega ^2 - 4 \omega + 5)} 
\Bigl[
\left(4 \omega ^2-4 \omega +5\right) 
\Bigl(
(2 \omega -1) ({\tilde x{}^1} {\tau})^{2 \omega }
+\gamma({\tilde x{}^2}^2  +{\tilde x{}^3}^2 )
\Bigr)
$$
$$
+
2 \gamma  (2 \omega -1) {\delta} 
\Bigl(
{\tilde x{}^2}^2+{\tilde x{}^2} {\tilde x{}^3} (2 \omega -1)-{\tilde x{}^3}^2
 \Bigr)
\sin \left(\theta -2\log \left({\tilde x{}^1} {\tau}\right)\right)
$$
\begin{equation}
+\gamma  (2 \omega -1) {\delta} 
\Bigl(
{\tilde x{}^2}^2 (2 \omega -1)-4 {\tilde x{}^2} {\tilde x{}^3}+{\tilde x{}^3}^2 (1-2 \omega )
\Bigr)
\cos \left(\theta -2\log \left({\tilde x{}^1} {\tau}\right)\right)
\Bigr]
\label{ToSynchr1}
, \end{equation}
$$
x^2 = 
-\frac{
\gamma  {\tau} ({\tilde x{}^1} {\tau})^{-2 \omega } 
}{(2 \omega -1 ) (4 \omega ^2 - 4 \omega + 5)}
\Bigl[
{\tilde x{}^2} \left(4 \omega ^2-4 \omega +5\right)
$$
$$
+
(2 \omega -1) {\delta} (2 {\tilde x{}^2}+{\tilde x{}^3} (2 \omega -1)) \sin \left(\theta -2\log \left({\tilde x{}^1} {\tau}\right)\right)
$$
\begin{equation}
+(2 \omega -1) {\delta} ({\tilde x{}^2} (2 \omega -1)-2 {\tilde x{}^3}) \cos \left(\theta -2\log \left({\tilde x{}^1} {\tau}\right)\right)
\Bigr]
\label{ToSynchr2}
, \end{equation}
$$
x^3 = 
\frac{\gamma  {\tau} ({\tilde x{}^1} {\tau})^{-2 \omega } 
}{(2 \omega -1 ) (4 \omega ^2 - 4 \omega + 5)} 
\Bigl[
{\tilde x{}^3} \left(-4 \omega ^2+4 \omega -5\right)
$$
$$
-(2 \omega -1) {\delta} ({\tilde x{}^2} (2 \omega -1)-2 {\tilde x{}^3}) \sin \left(\theta -2\log \left({\tilde x{}^1} {\tau}\right)\right)
$$
\begin{equation}
+(2 \omega -1) {\delta} (2 {\tilde x{}^2}+{\tilde x{}^3} (2 \omega -1)) \cos \left(\theta -2\log \left({\tilde x{}^1} {\tau}\right)\right)
\Bigr]
\label{ToSynchr3}
. \end{equation}

The chosen synchronous reference system is connected with the observer on the base geodesic, which in the synchronous reference system will have constant spatial coordinates. Four-velocity components of a freely falling observer on a base geodesic
(\ref{uPrivilUp0})-(\ref{uPrivilUp3}) are converted to the following form:
\begin{equation}
\tilde u^\alpha=\Bigl\{1,0,0,0\Bigr\}
. \end{equation}
Thus, the observer is at rest on the base geodesic relative to the chosen synchronous frame of reference.

Now, by applying the transformations (\ref{ToSynchr0})-(\ref{ToSynchr3}), we can write down the form of the gravitational wave metric in the synchronous frame of reference:
\begin{equation}
ds^2=d \tau^2-dl^2=d \tau^2+\tilde g{}_{{i}{j}}(\tau, {\tilde x{}^{k}})
\,d{\tilde x{}^{i}}d{\tilde x{}^{j}}
,\qquad
i,j,k=1,2,3;
\end{equation}
where $\tau$ is the time variable (the observer's proper time on the base geodesic), $dl$ is the spatial distance element, ${\tilde x{}^{k}}$ are the spatial coordinates.

The components of the gravitational wave metric (\ref{Metric}) in the synchronous frame take the following form ($\omega\ne 1/2$):
\begin{equation}
\label{metricSynchr1k}
\tilde g{}^{00} = 1 
,\qquad
\tilde g{}^{01} =
\tilde g{}^{02} = 
\tilde g{}^{03} = 0 
, \end{equation}
\begin{equation}
\tilde g{}^{1k} = -\frac{{\tilde x{}^1} {\tilde x{}^k}}{{\tau}^2} 
, \end{equation}
$$
\tilde g{}^{22} = 
\frac{
(1-2 \omega )^2 \left(\omega ^2-\omega -1\right) ({\tau} {\tilde x{}^1})^{2 \omega } 
}{5 \gamma  {\tau}^2}
\biggl[
-4 \omega ^2+4 \omega -5
$$
$$
\mbox{}
+
(2\omega+1)(2\omega-3)
{\delta} 
\cos \Bigl(\theta -2 \log ({\tau} {\tilde x{}^1})\Bigr)
$$
\begin{equation}
\mbox{}
+4 (2 \omega -1) {\delta} \sin \Bigl(\theta -2 \log ({\tau} {\tilde x{}^1})\Bigr)
\biggr]
-\frac{{\tilde x{}^2}^2}{{\tau}^2} 
\label{metricSynchr22}
, \end{equation}
$$
\tilde g{}^{23} = 
\frac{
(1-2 \omega )^2 \left(\omega ^2-\omega -1\right) {\delta} ({\tau} {\tilde x{}^1})^{2 \omega } 
}{5 \gamma  {\tau}^2}
\biggl[
4(1-2 \omega ) \cos \Bigl(\theta -2 \log ({\tau} {\tilde x{}^1})\Bigr)
$$
\begin{equation}
\mbox{}
+
(2\omega+1)(2\omega-3)
\sin \Bigl(\theta -2 \log ({\tau} {\tilde x{}^1})\Bigr)
\biggr]
-\frac{{\tilde x{}^2} {\tilde x{}^3}}{{\tau}^2} 
\label{metricSynchr23}
, \end{equation}
$$
\tilde g{}^{33} = 
-\frac{
(1-2 \omega )^2 \left(\omega ^2-\omega -1\right) ({\tau} {\tilde x{}^1})^{2 \omega } 
}{5 \gamma  {\tau}^2}
\biggl[
4 \omega ^2-4 \omega +5
$$
$$
\mbox{}
+
(2\omega+1)(2\omega-3)
{\delta} \cos \Bigl(\theta -2 \log ({\tau} {\tilde x{}^1})\Bigr)
$$
\begin{equation}
\mbox{}
+4 (2 \omega -1) {\delta} \sin \Bigl(\theta -2 \log ({\tau} {\tilde x{}^1})\Bigr)
\biggr]
-\frac{{\tilde x{}^3}^2}{{\tau}^2} 
\label{metricSynchr33}
, \end{equation}
where $\omega$, $\gamma$ and $\theta$ are independent parameters of the gravitational wave model, and the constant ${\delta}$ is given by the relation (\ref{deltaOmega}).

Converting the components of the deviation vector $\eta{}^\alpha (\tau)$ from the preferred coordinate system (\ref{DeviationSolutionPriv0})-(\ref{DeviationSolutionPriv3}) to the synchronous reference system gives the following result:
\begin{equation}
\tilde \eta{}^0= 0 
\label{DeviationSolutionSynchr0}
,\end{equation}
\begin{equation}
\tilde \eta{}^1 (\tau) = {\rho_1}-\frac{{\lambda_1} {\Omega}}{{\tau}} 
\label{DeviationSolutionSynchr1}
,\end{equation}
$$
\tilde \eta{}^2 (\tau) = 
\frac{(1-2 \omega ) \left(1+\omega -\omega ^2\right) {\lambda_1}{\vartheta_2} ({\lambda_1} {\tau})^{2 \omega-1 } 
}{5 \gamma  }
\Bigl[ 2 (1-2 \omega) {\delta} \sin \Bigl(\theta -2 \log ({\lambda_1} {\tau})\Bigr)
$$
$$
\mbox{}
-(1-2 \omega )^2 {\delta} \cos \Bigl(\theta -2 \log ({\lambda_1} {\tau})\Bigr)+4 \omega ^2-4 \omega +5\Bigr]
$$
$$
\mbox{}
-\frac{(1-2 \omega )^2 \left(1+\omega -\omega ^2\right) {\delta} {\lambda_1}{\vartheta_3} ({\lambda_1} {\tau})^{2 \omega-1 } 
}{5 \gamma }
\Bigl[
2 \cos \Bigl(\theta -2 \log ({\lambda_1} {\tau})\Bigr)
$$
\begin{equation}
\mbox{}
+(1-2 \omega ) \sin \Bigl(\theta -2 \log ({\lambda_1} {\tau})\Bigr)
\Bigr]
-\frac{{\lambda_2} {\Omega}}{{\tau}}+{\rho_2} 
\label{DeviationSolutionSynchr2}
,\end{equation}
$$
\tilde \eta{}^3 (\tau) =
 -\frac{(1-2 \omega )^2 \left(1+\omega -\omega ^2\right) 
 {\delta}{\lambda_1} {\vartheta_2} ({\lambda_1} {\tau})^{2 \omega -1} 
  }{5 \gamma }
 \Bigl[
 2 \cos \Bigl(\theta -2 \log ({\lambda_1} {\tau})\Bigr)
 $$
$$
\mbox{}
 + (1-2 \omega ) \sin \Bigl(\theta -2 \log ({\lambda_1} {\tau})\Bigr)
 \Bigr]
 $$
$$
\mbox{}
 +\frac{(1-2 \omega ) \left(1+\omega -\omega ^2\right) {\lambda_1}{\vartheta_3} ({\lambda_1} {\tau})^{2 \omega -1} 
}{5 \gamma }
 \Bigl[
 2 (2 \omega -1) {\delta} \sin \Bigl(\theta -2 \log ({\lambda_1} {\tau})\Bigr)
 $$
\begin{equation}
\mbox{}
 +(1-2 \omega )^2 {\delta} \cos \Bigl(\theta -2 \log ({\lambda_1} {\tau})\Bigr)+4 \omega ^2-4 \omega +5
 \Bigr]
 -\frac{{\lambda_3} {\Omega}}{{\tau}}+{\rho_3} 
 \label{DeviationSolutionSynchr3}
.\end{equation}
Here
the parameters $\omega$, $\gamma$ and $\theta$ determine the gravitational wave model,
the constants $\lambda_k$, $\rho_k$ and $\vartheta_k$ are given by the initial or boundary conditions for the velocities and mutual positions of particles on neighboring geodesics,
the auxiliary constant $\Omega$ is determined by the relation (\ref{constOmega}), the constant $\delta$ is determined by the relation (\ref{deltaOmega}).

Let us now find the form of tidal accelerations $\tilde A{}^\alpha$ of a gravitational wave (\ref{Metric})
in the synchronous reference system:
\begin{equation}
\tilde A{}^\alpha =\frac{ D^2\tilde\eta{}^\alpha}{d^2\tau}
 , \end{equation}
\begin{equation}
\tilde A{}^0= \tilde A{}^1= 0 
 , \end{equation}
$$
\tilde A{}^2= 
\frac{(2\omega -1)(\omega ^2-\omega -1) {\delta} {\vartheta_2} ({\lambda_1} {\tau})^{2 \omega } 
}{\gamma  {\tau}^3}
\Bigl[
2 \cos \Bigl(\theta -2 \log ({\lambda_1} {\tau})\Bigr)
$$
$$
\mbox{}
+(1-2 \omega ) \sin \Bigl(\theta -2 \log ({\lambda_1} {\tau})\Bigr)
\Bigr]
$$
$$
\mbox{}
+\frac{(2\omega -1)(\omega ^2-\omega -1){\delta} {\vartheta_3} ({\lambda_1} {\tau})^{2 \omega } 
}{\gamma  {\tau}^3}
\Bigl[
2 \sin \Bigl(\theta -2 \log ({\lambda_1} {\tau})\Bigr)
$$
$$
\mbox{}
+(2 \omega -1) \cos \Bigl(\theta -2 \log ({\lambda_1} {\tau})\Bigr)
\Bigr]
$$
$$
\mbox{}
+\frac{{R_2} \left(\omega ^2-\omega -1\right) {\delta} 
}{{\lambda_1} {\tau}^2}
\Bigl[
(1-2 \omega ) \sin \Bigl(\theta -2 \log ({\lambda_1} {\tau})\Bigr)+2 \cos \Bigl(\theta -2 \log ({\lambda_1} {\tau})\Bigr)
\Bigr]
$$
\begin{equation}
\mbox{}
+\frac{{R_3} 
}{{\lambda_1} {\tau}^2} 
\Bigl[
2 \left(\omega ^2-\omega -1\right) {\delta} \sin \Bigl(\theta -2 \log ({\lambda_1} {\tau})\Bigr)
$$
$$
\mbox{}
+(2\omega -1)(\omega ^2-\omega -1) {\delta} \cos \Bigl(\theta -2 \log ({\lambda_1} {\tau})\Bigr)+\omega  (2\omega-1)(1-\omega)
\Bigr]
 , \end{equation}
$$
\tilde A{}^3= \frac{(2\omega -1)(\omega ^2-\omega -1){\delta} {\vartheta_2} ({\lambda_1} {\tau})^{2 \omega } 
}{\gamma  {\tau}^3}
\Bigl[
2 \sin \Bigl(\theta -2 \log ({\lambda_1} {\tau})\Bigr)
$$
$$
\mbox{}
+(2 \omega -1) \cos \Bigl(\theta -2 \log ({\lambda_1} {\tau})\Bigr)
\Bigr]
$$
$$
\mbox{}
-\frac{(2\omega -1)(\omega ^2-\omega -1) {\delta} {\vartheta_3} ({\lambda_1} {\tau})^{2 \omega } 
}{\gamma  {\tau}^3}
\Bigl[
2 \cos \Bigl(\theta -2 \log ({\lambda_1} {\tau})\Bigr)
$$
$$
\mbox{}
+
(1-2 \omega ) \sin \Bigl(\theta -2 \log ({\lambda_1} {\tau})\Bigr)
\Bigr]
$$
$$
\mbox{}
+\frac{{R_2} 
}{{\lambda_1} {\tau}^2}
\Bigl[
\omega  (2\omega-1)(\omega-1)
+2 \left(\omega ^2-\omega -1\right) {\delta} \sin \Bigl(\theta -2 \log ({\lambda_1} {\tau})\Bigr)
$$
$$
\mbox{}
+(2\omega -1)(\omega ^2-\omega -1) {\delta} \cos \Bigl(\theta -2 \log ({\lambda_1} {\tau})\Bigr)
\Bigr]
$$
$$
\mbox{}
-\frac{{R_3} \left(\omega ^2-\omega -1\right) {\delta} 
}{{\lambda_1} {\tau}^2} 
\Bigl[
2 \cos \Bigl(\theta -2 \log ({\lambda_1} {\tau})\Bigr)
$$
\begin{equation}
\mbox{}
+(1-2 \omega ) \sin \Bigl(\theta -2 \log ({\lambda_1} {\tau})\Bigr)
\Bigr]
 , \end{equation}
where the parameters $\omega$, $\gamma$ and $\theta$ determine the characteristics of the gravitational wave, the parameters $\lambda_k$ are determined 
by the initial or boundary values of the particle velosity on the base geodesic,
the parameters ${\rho_k}$ and ${\vartheta_k}$ are determined 
by the initial or boundary values of the velosity
and relative positions of the particle on the adjacent geodesic, auxiliary constants $R_2$ and $R_3$ are defined by the relations (\ref{R23}).

In the synchronous reference frame used, the components of the gravitational wave tidal acceleration are not equal to zero only in the plane of variables ${\tilde x{}^{2}}$, ${\tilde x{}^{3}}$. The gravitational wave propagates along the space variable ${\tilde x{}^{1}}$.

\twocolumn

\section{A special case of a wave at $\omega=1/2$ in a privileged coordinate system}

When integrating the Hamilton-Jacobi equation of test particles for the Shapovalov type III space in the Bianchi type VII cosmological model for the metric (\ref{Metric}), a special case arises when
$\omega=1/2$. In this case, the wave metric acquires the following form in the privileged coordinate system
$$
{ds}^2=
2dx^0dx^1
-
\frac{
{x^0}
}{\gamma ^2-\alpha ^2-\beta ^2}
\Bigl[
$$
$$
\left(
\alpha  \cos{\bigl(2\log{x^0}\bigr)}+\beta  \sin{\bigl(2\log{x^0}\bigr)}-\gamma 
\right)
{dx^2}^2
$$
$$
+
\left(
\beta  \cos{\bigl(2\log{x^0}\bigr)}-\alpha  \sin{\bigl(2\log{x^0}\bigr)}
\right)
{dx^2}{dx^3}
$$
\begin{equation}
+
\left(
-\alpha  \cos{\bigl(2\log{x^0}\bigr)}-\beta  \sin{\bigl(2\log{x^0}\bigr)}-\gamma 
\right)
{dx^3}^2
\Bigr]
,\end{equation}
\begin{equation}
{g=\det g_{ij}=}-\frac{{x^0}^2}{\gamma ^2-\alpha ^2-\beta ^2}
.\end{equation}
The Einstein vacuum equations (\ref{EinsteinEqs}) in this case lead to a condition on the constants $\alpha$, $\beta$ and $\gamma$ of the form
\begin{equation}
\gamma^2 = 5 \left(\alpha ^2+\beta^2\right)
\label{constgamma}
,\end{equation}
which allows us to introduce a two-parameter representation ($\gamma$, $\theta$) using the constant angular parameter $\theta$ instead of the parameters $\alpha$ and $\beta$:
\begin{equation}
\label{GammaTheta}
\alpha=\frac{\gamma\sin{\theta}}{\sqrt{5}}
,\qquad
\beta=\frac{\gamma\cos{\theta}}{\sqrt{5}}
.\end{equation}
The metric signature $(+,-,-,-)$ leads to a negative value of the constant $\gamma$:
\begin{equation}
\gamma = -\sqrt{5 \left(\alpha ^2+\beta^2\right)} <0
,\end{equation}
Integrating the Hamilton-Jacobi equation of a test particle for the special case $\omega=1/2$, we obtain the complete integral in the following form
$$
S=
\sum_{k=1}^{3}{\lambda_k}{x^k}
+
 \frac{
 1
 }{4 {\lambda_1}} 
 \biggl[
2 {x^0} -2 \gamma  \left({\lambda_2}^2+{\lambda_3}^2\right) \log{x^0}
 $$
 $$
 \mbox{}
 +
 \Bigl(
 \beta  \left({\lambda_2}^2-{\lambda_3}^2\right)-2 \alpha  {\lambda_2} {\lambda_3}
 \Bigr)
 \cos{\left( 2\log{{x^0}}\right)}
$$
\begin{equation}
 \mbox{}
+
 \Bigl(
 \alpha  \left({\lambda_3}^2-{\lambda_2}^2\right)-2 \beta  {\lambda_2} {\lambda_3}
 \Bigr)
  \sin{\left(2\log{x^0}\right)}
\biggr]
,\end{equation}
where $\lambda_k$ are constant parameters,
determined by the initial or boundary values of the velocity of the test particle.

From the equations of motion in the Hamilton-Jacobi formalism, we find the general form of the trajectories of motion of test particles in a gravitational wave in a privileged coordinate system:
\begin{equation}
x^0 (\tau)  = {\lambda_1}{\tau} 
,\end{equation}
$$
x^1 (\tau)  = 
-\frac{
1
}{4 {\lambda_1}^2} 
\Bigl[
$$
$$
\left(\alpha  \left({\lambda_2}^2-{\lambda_3}^2\right)+2 \beta  {\lambda_2} {\lambda_3}\right) \sin{\bigl(2\log{({\lambda_1}{\tau})}\bigr)}
$$
$$
+\left(2 \alpha  {\lambda_2} {\lambda_3}+\beta  \left({\lambda_3}^2-{\lambda_2}^2\right)\right) \cos{\bigl(2\log{({\lambda_1}{\tau})}\bigr)}
$$
\begin{equation}
+2 \gamma  \left({\lambda_2}^2+{\lambda_3}^2\right) \log{({\lambda_1}{\tau})}-2 {\lambda_1}{\tau}
\Bigr]
,\end{equation}
$$
x^2 (\tau)  = 
\frac{
1}{2 {\lambda_1}} 
\Bigl[
(\alpha  {\lambda_2}+\beta  {\lambda_3}) \sin{\bigl(2\log{({\lambda_1}{\tau})}\bigr)}
$$
\begin{equation}
+(\alpha  {\lambda_3}-\beta  {\lambda_2}) \cos{\bigl(2\log{({\lambda_1}{\tau})}\bigr)}+2 {\lambda_2} \gamma  \log{({\lambda_1}{\tau})}
\Bigr]
,\end{equation}
$$
x^3 (\tau)  = 
\frac{
1}{2 {\lambda_1}} 
\Bigl[
(\beta  {\lambda_2}-\alpha  {\lambda_3}) \sin{\bigl(2\log{({\lambda_1}{\tau})}\bigr)}
$$
\begin{equation}
+(\alpha  {\lambda_2}+\beta  {\lambda_3}) \cos{\bigl(2\log{({\lambda_1}{\tau})}\bigr)}+2 {\lambda_3} \gamma  \log{({\lambda_1}{\tau})}
\Bigr]
,\end{equation}
where $\tau$ is the proper time of the test particle.

The components of the 4-velocity vector $u^\alpha$ of test particles in a gravitational wave take the form:
\begin{equation}
u^{0} = {\lambda_1} 
, \end{equation}
$$
u^{1} (\tau)  = 
\frac{
1}{2 {\lambda_1} {x^0}} 
\Bigl[
{x^0}
-\gamma ({\lambda_2}^2 +{\lambda_3}^2 )
$$
$$
\mbox{}
+
\sin{\bigl(2\log{x^0}\bigr)} \left(2 \alpha  {\lambda_2} {\lambda_3}-\beta  {\lambda_2}^2+\beta  {\lambda_3}^2\right)
$$
\begin{equation}
\mbox{}
+\cos{\bigl(2\log{x^0}\bigr)} \left(\alpha  \left({\lambda_3}^2-{\lambda_2}^2\right)-2 \beta  {\lambda_2} {\lambda_3}\right)
\Bigr]
, \end{equation}
$$
u^{2} (\tau)  = 
\frac{
1}{{x^0}} 
\Bigl[
\sin{\bigl(2\log{x^0}\bigr)} (\beta  {\lambda_2}-\alpha  {\lambda_3})
$$
\begin{equation}
\mbox{}
+\cos{\bigl(2\log{x^0}\bigr)} (\alpha  {\lambda_2}+\beta  {\lambda_3})
+{\lambda_2} \gamma 
\Bigr]
, \end{equation}
$$
u^{3} (\tau)  = 
\frac{
1
}{{x^0}} 
\Bigl[
\cos{\bigl(2\log{x^0}\bigr)} (\beta  {\lambda_2}-\alpha  {\lambda_3})
$$
\begin{equation}
\mbox{}
-\sin{\bigl(2\log{x^0}\bigr)} (\alpha  {\lambda_2}+\beta  {\lambda_3})
+{\lambda_3} \gamma 
\Bigr]
, \end{equation}
where $x^0 = {\lambda_1}{\tau}$.

\onecolumn

Solving the equations (\ref{Deviation1})-(\ref{Deviation2}) for the deviation vector $\eta^\alpha (\tau)$ , we obtain its following form in the privileged coordinate system:
\begin{equation}
 \eta^0 (\tau)  = {\rho_1}{\tau}  - {\lambda_1} {\Omega} 
, \qquad
x^0 = {\lambda_1}{\tau} 
, \end{equation}
$$
 \eta^1 (\tau)  = 
 \frac{{R_2} 
 }{2 {\lambda_1}^3}
\Bigl[
 \sin{\bigl(2\log{x^0}\bigr)} 
 (\alpha  {\lambda_2}+\beta  {\lambda_3})
$$
$$
 +
 \cos \left(2\log{{x^0}} \right)
(\alpha  {\lambda_3}-\beta  {\lambda_2})
 +2 \gamma {\lambda_2}   \log ({x^0})
\Bigr]
 $$
 $$
 \mbox{}
 +
 \frac{{R_3} 
  }{2 {\lambda_1}^3}
\Bigl[
 \sin{\bigl(2\log{x^0}\bigr)} (\beta  {\lambda_2}-\alpha  {\lambda_3})
 $$
 $$
 +\cos{\bigl(2\log{x^0}\bigr)} (\alpha  {\lambda_2}+\beta  {\lambda_3})+2  \gamma  {\lambda_3} \log ({x^0})
\Bigr]
 $$
 $$
 +
 \frac{
1 }{2 {\lambda_1}^3 {x^0}} 
 \Bigl[
 \sin{\bigl(2\log{x^0}\bigr)} \left(-2 \alpha  {\lambda_2} {\lambda_3}+\beta  {\lambda_2}^2-\beta  {\lambda_3}^2\right) \left({\lambda_1}^2 {\Omega}-{\rho_1} {x^0}\right)
 $$
 $$
 +\cos{\bigl(2\log{x^0}\bigr)} \left(\alpha  \left({\lambda_2}^2-{\lambda_3}^2\right)+2 \beta  {\lambda_2} {\lambda_3}\right) \left({\lambda_1}^2 {\Omega}-{\rho_1} {x^0}\right)
 $$
\begin{equation}
 \mbox{}
 -{\rho_1} {x^0}^2
-
  {x^0}
\left(
{\lambda_1}^2 {\Omega} 
+\gamma  {\rho_1} 
 \left(
 {\lambda_2}^2 
+{\lambda_3}^2
 \right)
\right)
 +
 {\lambda_1}^2 \gamma  {\Omega}
 \left(
 {\lambda_2}^2  + {\lambda_3}^2 
 \right)
\Bigr]
 +{\vartheta_1} 
, \end{equation}

$$
 \eta^2 (\tau)  =
  -\frac{
\left({\lambda_1}^2 {\Omega}-{\rho_1} {x^0}\right) 
}{{\lambda_1}^2 {x^0}}
\Bigl[
\sin{\bigl(2\log{x^0}\bigr)} (\beta  {\lambda_2}-\alpha  {\lambda_3})
$$
$$
+\cos{\bigl(2\log{x^0}\bigr)} (\alpha  {\lambda_2}+\beta  {\lambda_3})
+{\lambda_2} \gamma 
\Bigr]
$$
$$
 +\frac{
 {R_2} 
  }{2 {\lambda_1}^2}
\Bigl[
 -\alpha  \sin{\bigl(2\log{x^0}\bigr)}+\beta  \cos{\bigl(2\log{x^0}\bigr)}-2 \gamma  \log ({x^0})
\Bigr]
$$
\begin{equation}
 \mbox{}
 -\frac{{R_3} 
}{2 {\lambda_1}^2}
\Bigl[
 \alpha  \cos{\bigl(2\log{x^0}\bigr)}+\beta  \sin{\bigl(2\log{x^0}\bigr)}
\Bigr]
 +{\vartheta_2} 
, \end{equation}

$$
 \eta^3 (\tau)  =
  -\frac{
  \left({\lambda_1}^2 {\Omega}-{\rho_1} {x^0}\right) 
}{{\lambda_1}^2 {x^0}}
\Bigl[
  -\sin{\bigl(2\log{x^0}\bigr)} (\alpha  {\lambda_2}+\beta  {\lambda_3})
$$
$$
+\cos{\bigl(2\log{x^0}\bigr)} (\beta  {\lambda_2}-\alpha  {\lambda_3})+{\lambda_3} \gamma 
\Bigr]
$$
$$
-\frac{
{R_3} 
}{2 {\lambda_1}^2}
\Bigl[
-\alpha  \sin{\bigl(2\log{x^0}\bigr)}+\beta  \cos{\bigl(2\log{x^0}\bigr)}+2 \gamma  \log ({x^0})
\Bigr]
$$
\begin{equation}
 \mbox{}
-\frac{
{R_2} 
}{2 {\lambda_1}^2}
\Bigl[
\alpha  \cos{\bigl(2\log{x^0}\bigr)}+\beta  \sin{\bigl(2\log{x^0}\bigr)}
\Bigr]
 +{\vartheta_3} 
, \end{equation}
where the constant parameters ${\rho_k}$ and ${\vartheta_k}$ are determined by the initial or boundary values of the relative positions and velocities of neighboring geodesics. Auxiliary constants ${R_2}$, ${R_3}$ and $\Omega$ are introduced for brevity and are determined by the relations (\ref{R23})-(\ref{constOmega}).

Having obtained the solution for the deviation vector, we can now write down the form of tidal accelerations $A^\alpha (\tau)$ of neighboring geodesics. Calculating the covariant derivative $A^\alpha =D^2\eta^\alpha/{d\tau}^2$, we obtain the tidal acceleration components in the preferred coordinate system in the form:
\begin{equation}
A^0 = 0
, \qquad
x^0 = {\lambda_1}{\tau} 
,\end{equation}
$$
A^1 (\tau)  = 
\frac{
{R_2} 
}{4 {\lambda_1} {x^0}^2}
\biggl[
10 {\lambda_2} \log ({x^0}) 
\left(\alpha  \cos{\bigl(2\log{x^0}\bigr)}+\beta  \sin{\bigl(2\log{x^0}\bigr)}\right)
$$
$$
-{\lambda_3} 
\left(10 \alpha  \log ({x^0}) \sin{\bigl(2\log{x^0}\bigr)}-10 \beta  \log ({x^0}) \cos{\bigl(2\log{x^0}\bigr)}+\gamma \right)
\biggr]
$$
$$
+
\frac{
{R_3} 
}{4 {\lambda_1} {x^0}^2}
\biggl[
{\lambda_2} 
\left(-10 \alpha  \log ({x^0}) \sin{\bigl(2\log{x^0}\bigr)}+10 \beta  \log ({x^0}) \cos{\bigl(2\log{x^0}\bigr)}+\gamma \right)
$$
$$
- 10 {\lambda_3} \log ({x^0}) 
\left(\alpha  \cos{\bigl(2\log{x^0}\bigr)}+\beta  \sin{\bigl(2\log{x^0}\bigr)} \right)
\biggr]
$$
$$
-
\frac{
{5 {\lambda_1}\vartheta_2} 
}{2 \gamma  {x^0}^2}
\biggl[
 {\lambda_2} 
\left(\alpha  \cos{\bigl(2\log{x^0}\bigr)}+\beta  \sin{\bigl(2\log{x^0}\bigr)}\right)
$$
$$
+ {\lambda_3}
\left(\beta  \cos{\bigl(2\log{x^0}\bigr)} -\alpha  \sin{\bigl(2\log{x^0}\bigr)}\right)
\biggr]
$$
$$
+
\frac{
{5 {\lambda_1}\vartheta_3} 
}{2 \gamma  {x^0}^2} 
\biggl[
 {\lambda_3}
\left(\alpha  \cos{\bigl(2\log{x^0}\bigr)} +\beta  \sin{\bigl(2\log{x^0}\bigr)}\right)
$$
\begin{equation}
-  {\lambda_2} 
\left(\beta  \cos{\bigl(2\log{x^0}\bigr)}-\alpha  \sin{\bigl(2\log{x^0}\bigr)} \right)
\biggr]
,\end{equation}
$$
A^2 (\tau) = 
\frac{5 {\lambda_1}^2 {\vartheta_2}}{2 \gamma  {x^0}^2}
 \left(\alpha  \cos{\bigl(2\log{x^0}\bigr)}+\beta  \sin{\bigl(2\log{x^0}\bigr)}\right)
$$
$$
+\frac{5 {\lambda_1}^2 {\vartheta_3} }{2 \gamma  {x^0}^2}
\left(\beta  \cos{\bigl(2\log{x^0}\bigr)}-\alpha  \sin{\bigl(2\log{x^0}\bigr)}\right)
$$
$$
-\frac{{R_3}}{4 {x^0}^2}
 \left(-10 \alpha  \log ({x^0}) \sin{\bigl(2\log{x^0}\bigr)} 
 +10 \beta  \log ({x^0}) \cos{\bigl(2\log{x^0}\bigr)}+\gamma \right)
$$
\begin{equation}
-\frac{5 {R_2} \log ({x^0}) }{2 {x^0}^2}
\left(\alpha  \cos{\bigl(2\log{x^0}\bigr)}+\beta  \sin{\bigl(2\log{x^0}\bigr)}\right)
,\end{equation}
$$
A^3  (\tau)  = 
\frac{5 {\lambda_1}^2 {\vartheta_2} }{2 \gamma  {x^0}^2}
\left(\beta  \cos{\bigl(2\log{x^0}\bigr)}-\alpha  \sin{\bigl(2\log{x^0}\bigr)}\right)
$$
$$
-\frac{5 {\lambda_1}^2 {\vartheta_3} }{2 \gamma  {x^0}^2}
\left(\alpha  \cos{\bigl(2\log{x^0}\bigr)}+\beta  \sin{\bigl(2\log{x^0}\bigr)}\right)
$$
$$
+\frac{{R_2} }{4 {x^0}^2}
\left(10 \alpha  \log ({x^0}) \sin{\bigl(2\log{x^0}\bigr)} 
-10 \beta  \log ({x^0}) \cos{\bigl(2\log{x^0}\bigr)}+\gamma \right)
$$
\begin{equation}
+\frac{5 {R_3} \log ({x^0}) 
}{2 {x^0}^2}
\left(\alpha  \cos{\bigl(2\log{x^0}\bigr)}+\beta  \sin{\bigl(2\log{x^0}\bigr)}\right)
,\end{equation}
where the constant parameters of the gravitational wave $\alpha$, $\beta$, $\gamma$ are related by the relation (\ref{GammaTheta}), the constants $\lambda_k$, $\rho_k$, ${\vartheta_k}$ are given by the initial conditions for velocities and relative positions of particles on neighboring geodesics, the constants ${R_2}$ and ${R_3}$ are determined by the relations
(\ref{R23}).

The obtained exact solution for tidal acceleration in a gravitational wave can be used to calculate various physical effects when a wave passes through a material medium.

\twocolumn

\section{Special case of a wave for $\omega=1/2$ 
in synchronous reference system}

To study the physical effects of a gravitational wave and the action of tidal accelerations, it is convenient to switch to a synchronous reference frame, where the observer is at rest on the base geodesic, time and spatial coordinates are separated, and the observer’s proper time on the base geodesic is a temporal variable, and time can be synchronized between different points in space. time.

The transition to a similar synchronous reference frame $\tilde x{}^\alpha$ with a known complete integral of the Hamilton-Jacobi equation of test particles is carried out according to the well-known \cite{LandauEng1} algorithm by transformation to spatial variables, which are kept constant on the trajectory of the basic test particle:
\begin{equation}
x^\alpha \to \tilde x{}^\alpha=\left(\tau,\lambda_1,\lambda_2,\lambda_3 \right)
,\end{equation}
In our case $(\omega=1/2)$ the coordinate transformation takes the form:
\begin{equation}
x^0 = {\tilde x{}^1}{\tau} 
,\end{equation}
$$
x^1 = 
-\frac{
1
}{4 {\tilde x{}^1}^2} 
\Bigl[
$$
$$
\left(\alpha  \left({\tilde x{}^2}^2-{\tilde x{}^3}^2\right)+2 \beta  {\tilde x{}^2} {\tilde x{}^3}\right) 
\sin{\bigl(2\log{({\tilde x{}^1}{\tau})}\bigr)}
$$
$$
+\left(2 \alpha  {\tilde x{}^2} {\tilde x{}^3}+\beta  \left({\tilde x{}^3}^2-{\tilde x{}^2}^2\right)\right) 
\cos{\bigl(2\log{({\tilde x{}^1}{\tau})}\bigr)}
$$
\begin{equation}
+2 \gamma  \left({\tilde x{}^2}^2+{\tilde x{}^3}^2\right) \log{({\tilde x{}^1}{\tau})}-2 {\tilde x{}^1}{\tau}
\Bigr]
,\end{equation}
$$
x^2 = 
\frac{
1}{2 {\tilde x{}^1}} 
\Bigl[
(\alpha  {\tilde x{}^2}+\beta  {\tilde x{}^3}) \sin{\bigl(2\log{({\tilde x{}^1}{\tau})}\bigr)}
$$
$$
+(\alpha  {\tilde x{}^3}-\beta  {\tilde x{}^2}) 
\cos{\bigl(2\log{({\tilde x{}^1}{\tau})}\bigr)}
$$
\begin{equation}
+2 {\tilde x{}^2} \gamma  \log{({\tilde x{}^1}{\tau})}
\Bigr]
,\end{equation}
$$
x^3 = 
\frac{
1}{2 {\tilde x{}^1}} 
\Bigl[
(\beta  {\tilde x{}^2}-\alpha  {\tilde x{}^3}) 
\sin{\bigl(2\log{({\tilde x{}^1}{\tau})}\bigr)}
$$
$$
+(\alpha  {\tilde x{}^2}+\beta  {\tilde x{}^3}) 
\cos{\bigl(2\log{({\tilde x{}^1}{\tau})}\bigr)}
$$
\begin{equation}
+2 {\tilde x{}^3} \gamma  \log{({\tilde x{}^1}{\tau})}
\Bigr]
,\end{equation}

Upon transition to the considered synchronous frame of reference, the components of the gravitational wave metric will take the following form:
\begin{equation}
{ds}^2={d\tau}^2-{dl}^2={d\tau}^2+\tilde g{}_{ij}(\tau,\tilde x{}^k)\,d\tilde x{}^id\tilde x{}^j
,\end{equation}
where ${dl}$ is the spatial distance element, which in the synchronous reference frame is a function of time $\tau$ and spatial variables $\tilde x{}^k$.

The four-velocity of the particle on the base geodesic in the chosen synchronous frame of reference, as expected, takes the following form:
\begin{equation}
\tilde u{}^\alpha=\Bigl\{ 1,0,0,0 \Bigr\}
,\end{equation}
those, in the selected synchronous frame of reference, the observer rests on the base geodesic.

The metric of a gravitational wave in the Bianchi type VII space in the considered synchronous reference frame in the case of $\omega=1/2$ will take the following form:
\begin{equation}
\tilde g{}^{00} = 1 
,\qquad
\tilde g{}^{01} = 
\tilde g{}^{02} = 
\tilde g{}^{03} = 0 
,\end{equation}
\begin{equation}
\tilde g{}^{1k} = -\frac{{\tilde x{}^1} {\tilde x{}^k}}{{\tau}^2} 
, \end{equation}
$$
\tilde g{}^{22} = 
-\frac{\left({\tilde x{}^2}\right)^2}{{\tau}^2} 
$$
$$
\mbox{}
+
\frac{
20\, {\tilde x{}^1} 
}{\gamma^2{\tau}  \Bigl( 1 -20  \log ^2({\tilde x{}^1}{\tau}) \Bigr)^2}
\biggl[
\gamma 
+
20 \gamma  \log ^2({\tilde x{}^1}{\tau})
$$
$$
+
\sin{\bigl(2\log{({\tilde x{}^1}{\tau})}\bigr)} 
\left(
20 \beta  \log ^2({\tilde x{}^1}{\tau})
-20 \alpha  \log{({\tilde x{}^1}{\tau})}-\beta 
\right)
$$
\begin{equation}
+\cos{\bigl(2\log{({\tilde x{}^1}{\tau})}\bigr)} 
\left(
20 \alpha  \log ^2({\tilde x{}^1}{\tau}) +20 \beta  \log{({\tilde x{}^1}{\tau})}
-\alpha
\right)
\biggr]
, \end{equation}
$$
\tilde g{}^{23} = 
-\frac{{\tilde x{}^2} {\tilde x{}^3}}{{\tau}^2} 
+
\frac{
20\, {\tilde x{}^1} 
}{\gamma^2{\tau} \Bigl( 1 -20  \log ^2({\tilde x{}^1}{\tau}) \Bigr)^2}
\biggl[
$$
$$
\Bigl(\alpha 
- 20
\log{({\tilde x{}^1}{\tau})}
\left(
 \alpha  \log{({\tilde x{}^1}{\tau})} + \beta  
 \right)
\Bigr)
\sin{\bigl(2\log{({\tilde x{}^1}{\tau})}\bigr)} 
$$
\begin{equation}
+\left(20 \beta  \log ^2({\tilde x{}^1}{\tau}) - 20 \alpha  \log{({\tilde x{}^1}{\tau})}-\beta \right)
\cos{\bigl(2\log{({\tilde x{}^1}{\tau})}\bigr)} 
\biggr]
, \end{equation}
$$
\tilde g{}^{33} = 
-\frac{\left({\tilde x{}^3}\right)^2}{{\tau}^2} 
$$
$$
\mbox{}
+
\frac{
20\, {\tilde x{}^1}
}{\gamma^2{\tau}  \Bigl( 1 -20   \log ^2({\tilde x{}^1}{\tau})\Bigr)^2}
\biggl[
\gamma 
\left(
1
+
20   \log ^2({\tilde x{}^1}{\tau})
\right)
$$
$$
\mbox{}
+
\cos{\bigl(2\log{({\tilde x{}^1}{\tau})}\bigr)} 
\left(
\alpha
 -20 \beta  \log{({\tilde x{}^1}{\tau})}
-20 \alpha  \log ^2({\tilde x{}^1}{\tau})
\right)
$$
\begin{equation}
\mbox{}
+
\sin{\bigl(2\log{({\tilde x{}^1}{\tau})}\bigr)} 
\left(
\beta 
+
20 \alpha  \log{({\tilde x{}^1}{\tau})}-20 \beta  \log ^2({\tilde x{}^1}{\tau})
\right)
\biggr]
, \end{equation}
where the observer's proper time on the base geodesic $\tau$ is the unified time of the synchronous frame of reference, and the variables ${\tilde x{}^k}$ are the spatial coordinates.

The resulting coordinate transformation also allows calculating the components of the geodesic deviation vector, and since the base geodesic in the used synchronous reference system has become a time line on which the spatial coordinates do not change, the deviation vector, up to a shift in the reference point, becomes simply the spatial position (radius vector) of the test particles on an adjacent geodesic.

The components of the 4-vector deviation $\tilde\eta{}^\alpha(\tau)$ in the synchronous frame of reference in the case of $\omega=1/2$ take the form:
\begin{equation}
\tilde\eta{}^0= 0 
, \end{equation}
\begin{equation}
\tilde\eta{}^1 (\tau) = -\frac{{\lambda_1} {\Omega}}{{\tau}} 
+{\rho_1}
, \end{equation}
$$
\tilde\eta{}^2 (\tau) = 
\frac{
10 {\lambda_1} {\vartheta_2} 
}{\gamma ^2 \left(20 \log ^2({\lambda_1}{\tau})-1\right)}
\Bigl[
-\alpha  \sin{\bigl(2\log{({\lambda_1}{\tau})}\bigr)}
$$
$$
+\beta  \cos{\bigl(2\log{({\lambda_1}{\tau})}\bigr)}+2 \gamma  \log ({\lambda_1}{\tau})
\Bigr]
$$
$$
\mbox{}
-\frac{
10 {\lambda_1} {\vartheta_3} 
}{\gamma ^2 \left(20 \log ^2({\lambda_1}{\tau})-1\right)}
\Bigl[
\alpha  \cos{\bigl(2\log{({\lambda_1}{\tau})}\bigr)}
$$
\begin{equation}
\mbox{}
+\beta  \sin{\bigl(2\log{({\lambda_1}{\tau})}\bigr)}
\Bigr]
-\frac{{\lambda_2} {\Omega}}{{\tau}}
+{\rho_2} 
, \end{equation}
$$
\tilde\eta{}^3 (\tau) =
\frac{
 10 {\lambda_1} {\vartheta_3} 
  }{\gamma ^2 \left(20 \log ^2({\lambda_1}{\tau})-1\right)}
\Bigl[
 \alpha  \sin{\bigl(2\log{({\lambda_1}{\tau})}\bigr)}
$$
$$
-\beta  \cos{\bigl(2\log{({\lambda_1}{\tau})}\bigr)}
 +2 \gamma  \log{({\lambda_1}{\tau})}
\Bigr]
$$
$$
\mbox{}
  -\frac{
 10 {\lambda_1} {\vartheta_2}  
  }{\gamma ^2 \left(20 \log ^2({\lambda_1}{\tau})-1\right)}
\Bigl[
 \alpha  \cos{\bigl(2\log{({\lambda_1}{\tau})}\bigr)}
 $$
\begin{equation}
\mbox{}
+\beta  \sin{\bigl(2\log{({\lambda_1}{\tau})}\bigr)}
\Bigr]
 -\frac{{\lambda_3} {\Omega}}{{\tau}}
 +{\rho_3} 
, \end{equation}
where the quantities $\lambda_k$ now determine the constant spatial coordinates of the observer on the base geodesic, $\tau$ is the time, the constant parameters ${\rho_k}$ determine the asymptotic values of the spatial components of the deviation vector
for large values of $\tau$.

\onecolumn

Let us find in the case of $\omega=1/2$ in the synchronous frame of reference
4-vector of tidal acceleration $\tilde A{}^\alpha(\tau) = D^2\tilde\eta{}^\alpha/{d\tau}^2$
in a gravitational wave:
\begin{equation}
\tilde A^0= 0 
,\qquad
\tilde A^1= 0 
, \end{equation}
$$
\tilde A^2 (\tau) = 
\frac{50 {\lambda_1} {\vartheta_2} \log{({\lambda_1}{\tau})} 
\left(\alpha  \cos{\bigl( 2\log{({\lambda_1}{\tau})}\bigr)}
+\beta  \sin{\bigl( 2\log{({\lambda_1}{\tau})}\bigr)}\right)}{\gamma ^2{\tau}^2 \left(20 \log ^2({\lambda_1}{\tau})-1\right)}
$$
$$
\mbox{}
-\frac{5 {\lambda_1} {\vartheta_3} 
\left(10 \alpha  \log ({\lambda_1} {\tau}) \sin{\bigl( 2\log{({\lambda_1}{\tau})}\bigr)}
-10 \beta  \log{({\lambda_1}{\tau})} \cos{\bigl( 2\log{({\lambda_1}{\tau})}\bigr)}-\gamma \right)
}{\gamma ^2 {\tau}^2 \left(20 \log ^2({\lambda_1}{\tau})-1\right)}
$$
$$
\mbox{}
-\frac{5 {R_3} 
}{2 {\lambda_1} \gamma {\tau}^2 \left(20 \log ^2({\lambda_1}{\tau})-1\right)} 
\Bigl[
-\alpha  \left(20 \log ^2({\lambda_1}{\tau})+1\right) 
\sin{\bigl( 2\log{({\lambda_1}{\tau})}\bigr)}
$$
$$
\mbox{}
+\beta  \left(20 \log ^2({\lambda_1}{\tau})+1\right) \cos{\bigl( 2\log{({\lambda_1}{\tau})}\bigr)}+4 \gamma  \log{({\lambda_1}{\tau})}
\Bigr]
$$
\begin{equation}
\mbox{}
-\frac{5 {R_2} 
\left(20 \log ^2({\lambda_1}{\tau})+1\right) 
\left[\alpha  \cos{\bigl( 2\log{({\lambda_1}{\tau})}\bigr)}
+\beta  \sin{\bigl( 2\log{({\lambda_1}{\tau})}\bigr)}\right]
}{2 {\lambda_1} \gamma {\tau}^2 
\left(20 \log ^2({\lambda_1}{\tau})-1\right)
}
, \end{equation}

$$
\tilde A^3 (\tau) = 
-\frac{50 {\lambda_1} {\vartheta_3} \log{({\lambda_1}{\tau})} 
\left[
\alpha \cos{\bigl( 2\log{({\lambda_1}{\tau})}\bigr)}
+\beta  \sin{\bigl( 2\log{({\lambda_1}{\tau})}\bigr)}
\right]
}{\gamma ^2 {\tau}^2 \left(20 \log ^2({\lambda_1}{\tau})-1\right)}
$$
$$
\mbox{}
-\frac{5 {\lambda_1} {\vartheta_2} 
\left[
10 \alpha  \log{({\lambda_1}{\tau})} 
\sin{\bigl( 2\log{({\lambda_1}{\tau})}\bigr)}-10 \beta  \log ({\lambda_1}{\tau}) \cos{\bigl( 2\log{({\lambda_1}{\tau})}\bigr)}+\gamma 
\right]
}{\gamma ^2{\tau}^2 \left(20 \log ^2({\lambda_1}{\tau})-1\right)}
$$
$$
\mbox{}
+\frac{5 {R_2} 
}{2 {\lambda_1} \gamma {\tau}^2 \left(20 \log ^2({\lambda_1}{\tau})-1\right)}
\Bigl[
\alpha  \left(20 \log ^2({\lambda_1}{\tau})+1\right) 
\sin{\bigl( 2\log{({\lambda_1}{\tau})}\bigr)}
$$
$$
\mbox{}
-\beta  \left(20 \log ^2({\lambda_1}{\tau})+1\right) \cos{\bigl( 2\log{({\lambda_1}{\tau})}\bigr)}+4 \gamma  \log{({\lambda_1}{\tau})}
\Bigr]
$$
\begin{equation}
\mbox{}
+\frac{5 {R_3} 
\left(20 \log ^2({\lambda_1}{\tau})+1\right) 
\left[
\alpha \cos{\bigl( 2\log{({\lambda_1}{\tau})}\bigr)}
+\beta  \sin{\bigl( 2\log{({\lambda_1}{\tau})}\bigr)}
\right]
}{2 {\lambda_1} \gamma {\tau}^2 
\left(20 \log ^2({\lambda_1}{\tau})-1\right)
} 
, \end{equation}
where $\tau$ is the time in the synchronous frame of reference, the constants $\lambda_k$ 
are the values of the spatial coordinates
$\tilde x{}^k$ of the observer on the base geodesic, the constants $\alpha$, $\beta$ and $\gamma$ are the parameters of a gravitational wave in type VII Bianchi spaces, related by an additional relation (\ref{constgamma}).

In the synchronous reference frame used, tidal accelerations appear only in the plane of variables $\tilde x{}^2$ and $\tilde x{}^3$, while the wave propagates along the coordinate $\tilde x{}^1$.

\section{On retarded potentials generated by a charge in a gravitational wave}

As an example of the application of the obtained exact solutions, consider the problem of retarded potentials.

In the synchronous reference frame we use, the observer is at rest on the base geodesic, and the motion of the test particle on the neighboring geodesic is carried out with tidal acceleration, so if this particle has a charge, we can formulate the problem of finding retarded potentials for a charge moving along a given trajectory in a gravitational wave with respect to resting observer in a synchronous frame of reference.

The exact solution of this problem requires the integration of Maxwell's equations in space-time with the metric (\ref{metricSynchr1k})-(\ref{metricSynchr33}), when the charge motion is given by the deviation vector (\ref{DeviationSolutionSynchr0})-(\ref{DeviationSolutionSynchr3}).

\twocolumn

The solution of such a problem in flat space-time is given, as is known, by the Lienard-Wiechert potentials $A^\alpha$ (in this section $A^\alpha$ is used to denote electromagnetic potentials), which can be written as \cite{LandauEng1}:
\begin{equation}
A^\alpha(\tau, \vec{r})=e\frac{u^\alpha}{R_{\beta} u^{\beta}}
\label{PotentialEq}
,\end{equation}
where $e$ is the charge of the particle, $u^\alpha$ is the 4-velocity of the charge at time $\tau'$,
$R^{\beta}=\{c(\tau-\tau'), \vec{r}-\vec{r}\,{}'\}$ is the difference of the observer's 4-vector at time 
$\tau$ and the 4-vector of the charge at the time $\tau'$, $c$ is the speed of light, $\vec{r}$ is the spatial radius vector of the charge, and the time $\tau'$ and the components of the radius vector charges $\vec{r}\,{}'$ are related to $\tau$ and $\vec{r}$ by the following relation:
\begin{equation}
R_\alpha R^\alpha=0
.\end{equation}

In the weak field approximation, when the space-time differs little from the Minkowski space, we can apply the Lienard-Wiechert potential formula for an approximate, qualitative study of the problem, but taking into account the specifics of the exact solution we obtained for the deviation vector $\eta^\alpha$.

In the weak field approximation, for the charge vector $x^\alpha(\tau)$ and the vector $R^\alpha$ in the synchronous reference frame, one can take the expressions
\begin{equation}
\tilde x{}^\alpha (\tau)=
\left\{c\tau, \, \vec{\eta}(\tau)+\vec{\lambda}\,\right\}
,\end{equation}
\begin{equation}
R^\alpha=\left\{
c(\tau-\tau'),-\vec{\eta}(\tau')
\right\}
,\end{equation}
\begin{equation}
\vec{\eta}=\left\{ 
\tilde\eta{}^1, \tilde\eta{}^2, \tilde\eta{}^3
\,\right\}
,\quad
\vec{\lambda}=\left\{ 
{\lambda_1}, {\lambda_2}, {\lambda_3}
\,\right\}
.\end{equation}

Then the retarded potentials of the charge in the gravitational wave at the observer's point on the base geodesic can be represented as (\ref{PotentialEq}), where $\tilde\eta{}^\alpha$ is the deviation vector (\ref{DeviationSolutionSynchr0})-(\ref{DeviationSolutionSynchr3}), and in the expression for the potentials $A^\alpha(\tau)$
the values on the right
 are taken at the moment time ${\tau'}$, which is related to the time $\tau$ in the synchronous reference frame by the relation
\begin{equation}
c^2 (\tau-\tau')^2 +\tilde g{}^{kl}
\,\tilde \eta{}_k \tilde \eta{}_l
\,
\Bigr\rvert_{x^\alpha=\{c\tau',\,\vec{\lambda} \} }=0
,\end{equation}
where $k,l=1,2,3$ and the gravitational wave metric $\tilde g{}^{\alpha\beta}$ in the synchronous reference frame is taken from the relations (\ref{metricSynchr1k})-(\ref{metricSynchr33}).

\section{Discussion}

The exact solutions of the geodesic deviation equation obtained in this work are the basis for calculating various physical effects in a gravitational wave, including
to calculate the radiation of charges moving with tidal acceleration in a gravitational wave.
These solutions make it possible to estimate the influence of primordial  gravitational waves on the parameters of the microwave background of the Universe. Even a “naive” direct substitution into the Lienard–Wiechert formulas of the solutions obtained in this work for the deviation vector gives, in the weak field approximation, a qualitative description of the radiation of charged plasma particles in the primordial gravitational wave.

Obtaining exact mathematical models of geodesic deviation in primordial  gravitational waves with different types of space-time symmetries also makes it possible to evaluate the influence of primordial  gravitational waves on the formation of the modern stage of homogeneous and isotropic space-time in order to understand the process of "isotropization" in the early stages of the Universe (including for describing violations of the homogeneity and isotropy of the observed microwave electromagnetic background of the Universe), which is important for constructing theoretical models of the early Universe.

Exact models of primordial  gravitational waves for different types of space-time symmetries will make it possible to reveal those parameters of the electromagnetic microwave background that can characterize the presence of certain types of spatial homogeneity symmetries in the initial stages of the Universe, which is significant for constructing theoretical models of the early Universe.

The approach proposed in this paper for obtaining solutions to geodesic deviation equations in models of primordial gravitational waves for Universes with Bianchi symmetries is applicable both in Einstein's theory of gravity and in modified theories of gravity. This allows, using exact solutions for the deviation vector in various models and modified theories of gravitation, to obtain different options for the influence of primary plasma radiation and to find out the differences that may arise in the parameters of the microwave electromagnetic background of the Universe at the present stage under the influence of primordial gravitational waves in various models and modified theories of gravity.

\section{Conclusion}

An exact solution of the geodesic deviation equations in a gravitational wave is obtained for Bianchi type VII cosmological models in Einstein's theory of gravitation. The models of gravitational waves under consideration refer to Shapovalov wave spaces of type III, which made it possible to obtain a complete integral for the Hamilton-Jacobi equation of test particles in a privileged coordinate system. The presence of the complete integral of the Hamilton-Jacobi equation made it possible, in turn, to obtain both the trajectories of motion of test particles and the solution of the geodesic deviation equations in the models under consideration.

The obtained exact solutions of the geodesic deviation equation are presented both in the privileged coordinate system and in the synchronous (laboratory) coordinate system, relative to which the freely falling observer is at rest. The transition to a synchronous coordinate system is based on the use of the trajectories of motion of test particles obtained in the work.
An explicit form of metrics, geodesic deviation vectors, and an explicit form of tidal force acceleration in a gravitational wave are obtained in this work both for a privileged coordinate system and for a synchronous coordinate system.

The approach presented in the paper can be used both in the general theory of relativity and in modified theories of gravity. The exact models obtained describe the primordial gravitational waves of the Universe and can be used to calculate the secondary physical effects that arise during the passage of a wave, as well as to debug approximate and numerical methods of detection and comparative analysis of the characteristics of gravitational waves in terms of their effect on test particles and charges.




\section*{Acknowledgments}

The authors thank the administration of the Tomsk State Pedagogical University for the technical support of the scientific project. 
The study was supported by the Russian Science Foundation, grant \mbox{No. 22-21-00265},
\url{https://rscf.ru/project/22-21-00265/}

\section*{Statements and Declarations}

\subsection*{Data availability statement}

All necessary data and references to external sources are contained in the text of the manuscript. 
All information sources used in the work are publicly available and refer to open publications in scientific journals and textbooks.

\subsection*{Compliance with Ethical Standards}

The authors declare no conflict of interest.

\subsection*{Competing Interests and Funding}
The study was supported by the Russian Science Foundation, 
grant \mbox{No. 22-21-00265}, 
\url{https://rscf.ru/project/22-21-00265/}





\end{document}